\documentclass{jpsj2}

\usepackage{subfigure}
\usepackage{color}

\title{Approach to equilbrium in nano-scale systems at finite temperature}

\author{
\textsc{Fengping Jin}$^{1}$\thanks{E-mail: f.jin@rug.nl},
\textsc{Hans De Raedt}$^{1}$\thanks{E-mail: h.a.de.raedt@rug.nl},
\textsc{Shengjun Yuan }$^{2}$\thanks{E-mail: s.yuan@science.ru.nl},
\textsc{Mikhail I. Katsnelson}$^{2}$\thanks{E-mail: m.i.katsnelson@science.ru.nl},
\textsc{Seiji Miyashita}$^{3}$\thanks{E-mail: miya@spin.phys.s.u-tokyo.ac.jp}
and
\textsc{Kristel Michielsen}$^{4}$\thanks{E-mail: k.michielsen@fz-juelich.de}
}%

\inst{
$^{1}$Department of Applied Physics, Zernike Institute for Advanced Materials,\\
University of Groningen, Nijenborgh 4, NL-9747AG Groningen, The Netherlands\\
$^{2}$Institute of Molecules and Materials, Radboud University of Nijmegen,\\
NL-6525ED Nijmegen, The Netherlands \\
$^{3}$Department of Physics, Graduate School of Science,
The University of Tokyo,\\ 7-3-1 Hongo, Bunkyo-Ku, Tokyo 113-8656
and \\
CREST, JST, 4-1-8 Honcho Kawaguchi, Saitama 332-0012, Japan\\
$^{4}$Institute for Advanced Simulation, J\"ulich Supercomputing Centre,\\
Research Centre J\"ulich, D-52425 J\"ulich, Germany
}%

\abst{%
We study the time evolution of the reduced density matrix
of a system of spin-1/2 particles interacting with an environment of spin-1/2 particles.
The initial state of the composite system is taken to be a product state of a pure state of the system and a pure
state of the environment.
The latter pure state is prepared such that it represents the environment at a given finite temperature
in the canonical ensemble.
The state of the composite system evolves according to the time-dependent Schr{\"{o}}dinger equation,
the interaction creating entanglement between the system and the environment.
It is shown that independent of the strength of the interaction and the initial
temperature of the environment, all the eigenvalues of the reduced density matrix converge to
their stationary values, implying that also the entropy of the system relaxes to a stationary value.
We demonstrate that the difference between the canonical density matrix and the reduced density matrix
in the stationary state increases as the initial temperature of the environment decreases.
As our numerical simulations are necessarily restricted to a modest number of spin-1/2 particles ($<36$),
but do not rely on time-averaging of observables nor on the assumption that the coupling between system and environment is weak,
they suggest that the stationary state of the system directly follows from the time evolution of a pure state of the composite system,
even if the size of the latter cannot be regarded as being close to the thermodynamic limit.
}

\kword{Quantum Statistical Mechanics, Canonical Ensemble, Time-dependent Schr{\"o}dinger Equation, Thermalization, Decoherence}

\begin{document}
\maketitle

\section{Introduction}\label{Introduction}

Statistical mechanics is one of the cornerstones of modern physics~\cite{KUBO85,GREI97}
but its foundations are still subject of much
research~\cite{SCHR52,BOCC59,JENS85,SAIT96,TASA98,GEMM01,GEMM03,GOLD06,POPE06,GEMM06,CAZA06,REIM07,RIGO07,MERK07,ECKS08,CRAM08a,CRAM08b,LIND09,CHO10,GENW10,GOLD10}.
Some fundamental questions, such as how the canonical distribution emerges from the interaction between a system $S$
and its environment $E$, have only been partially resolved.
Answers to this question are important since the canonical ensemble is widely used to calculate the thermodynamic
quantities of a system at a given temperature.

As well for classical as quantum systems it is well-known that if the interaction between a system $S$ and its much larger environment $E$
having a large number of degrees of freedom and a dense distribution of energy levels, is weak, the system $S$ is described by a canonical
ensemble when the composite system $S+E$ is described by the microcanonical ensemble with a given total energy.
Usually, the derivation of the canonical distribution is discussed under the hypothesis that each state in the microcanonical ensemble
has equal probability~\cite{KUBO85,GREI97}.
In the case of quantum systems, it has recently been shown that the microcanonical mixed state for the composite system $S+E$
is not a required starting point for the system $S$ to be described by a canonical ensemble, but that $S+E$ being initially in a
randomly picked pure state with small energy fluctuations is sufficient~\cite{TASA98,GOLD06,POPE06,REIM07}.
The characteristic that even if the state of the composite quantum system $S+E$ corresponds to a single wave function only,
the reduced density matrix of $S$ is canonical for the overwhelming majority of wave functions in the subspace corresponding
to the energy interval encompassed by the microcanonical ensemble, is referred to as {\it canonical typicality} after Ref.~\cite{GOLD06}.
Not explicitly mentioning canonical typicality, this characteristic had already been used to calculate the density of states (DOS) of quantum
many-body systems~\cite{VRIE93,HAMS00}.
More recently it has been shown that canonical typicality has a classical counterpart: For {\it typical} probability distributions
defined on an energy shell of the classical composite system $S+E$, i.e. not necessarily microcanonical distributions, the marginal
probability distribution corresponding to the system $S$ exhibits the canonical form~\cite{PLAS08}.

In this paper we focus on the equilibration obtained from the dynamics of relatively small closed quantum systems (containing less than
36 spin-1/2 particles) that we compose from a small system $S$ and a much larger but still relatively small (containing less than 32 spin-1/2 particles)
environment $E$. We use general quantum spin-1/2 Hamiltonians to describe $S$ and $E$ and we do not put any restriction
on their energy spectra.
We study the conditions under which the stationary state of the system $S$ is represented by a canonical ensemble density matrix,
although the standard conditions, such as the environment $E$ being very large and the coupling between $S$ and $E$ being weak, are not
(necessarily) fulfilled.
The approach we use is to first solve the time-dependent Schr{\"{o}}dinger equation (TDSE) of the composite system $S+E$ of spin-1/2 particles
numerically and then analyze the behavior of the reduced density matrix of the system $S$, obtained by tracing out the environment $E$,
which was initially prepared in a randomly picked ``typical'' pure state with a given temperature.

Earlier work adopting this approach showed that a nanoscale environment prepared in a uniform random superposition of all states
(corresponding to an environment at infinite temperature)
can drive the system to the state with a canonical distribution~\cite{YUAN09}
and elucidated the effect of frustration and connectivity
on decoherence and relaxation processes~\cite{YUAN06,YUAN07,YUAN08}.
In this paper, we extend the approach to
study the decoherence and relaxation properties of nanoscale magnets embedded
in a nanoscale magnetic environment at finite temperature.

Our simulation results show that, independent of the strength of the interaction between $S$ and $E$ and the initial temperature of $E$,
$S$ evolves to a stationary state of which the properties strongly depend on the initial temperature of $E$.
This equilibration is remarkable given the relative small size of $E$, since usually
in equilibration studies the hypothesis of having a large environment is essential~\cite{LIND09,REIM07}.
We show that for sufficiently large initial temperatures of $E$,
the stationary state of $S$ is represented by a canonical ensemble density matrix at some finite effective temperature.
For decreasing temperatures, the reduced density matrix of $S$ deviates from the canonical density matrix.
The deviation increases for decreasing values of the interaction strength between $S$ and $E$.

The paper is organized as follows.
In Section 2, we discuss the model, define the quantities of interest and
summarize the essentials of the simulation method used.
The simulation results are presented in Section 3.
A discussion and conclusion is given in Section 4.

\section{Generalities}

In general, the state of a closed quantum system is described by a density matrix~\cite{NEUM55,BALL03}.
The canonical ensemble is characterized by a density matrix that is diagonal with respect
to the eigenstates of the Hamiltonian $H$, the diagonal elements taking the form
$\exp(-\beta E_{i})$ where $\beta=1/k_BT$ is proportional to the inverse temperature
($k_B$ is Boltzmann's constant and is taken to be one in this paper)
and the $E_{i}$'s denote the eigenenergies of $H$~\cite{KUBO85,GREI97}.

The time evolution of a closed quantum system is governed by the TDSE~\cite{NEUM55,BALL03}.
If the initial density matrix of an isolated quantum system is non-diagonal then, according to the time evolution
dictated by the TDSE, it remains non-diagonal and the quantum system never approaches the thermal
equilibrium state with the canonical distribution.
Therefore, in order to equilibrate the system $S$, it is necessary to have the
system $S$ interact with an environment $E$, also called a heat bath.
Thus, the Hamiltonian of the composite system $S+E$ takes
the form $H=H_{S}+H_{E}+H_{SE}$, where $H_S$ and $H_E$ are the system and environment Hamiltonian, respectively and $H_{SE}$
describes the interaction between the system and environment.

\subsection{Model}

To study the evolution to the canonical ensemble state in detail,
we consider a general quantum spin-1/2 model defined by the Hamiltonian
$H=H_{S}+H_{E}+H_{SE}$ where
\begin{eqnarray}
H_{S} &=&-\sum_{i=1}^{n_{S}-1}\sum_{j=i+1}^{n_{S}}\sum_{\alpha
=x.y,z}J_{i,j}^{\alpha }S_{i}^{\alpha }S_{j}^{\alpha }, \label{HAMS} \\ 
H_{E} &=&-\sum_{i=1}^{n_E-1}\sum_{j=i+1}^{n_E}\sum_{\alpha =x,y,z}\Omega
_{i,j}^{\alpha }I_{i}^{\alpha }I_{j}^{\alpha },  \label{HAME}\\ 
H_{SE} &=&-\sum_{i=1}^{n_{S}}\sum_{j=1}^{n_E}\sum_{\alpha =x,y,z}\Delta
_{i,j}^{\alpha }S_{i}^{\alpha }I_{j}^{\alpha }.  \label{HAMSE}
\end{eqnarray}%
Here $S$ and $I$ denote the spin-1/2 operators
of the system and environment, respectively (we use units such that
$\hbar$ and $k_B$ are one).
The total number of spins in the system and environment are denoted by $n_S$ and $n_E$, respectively.

The spins of the system are arranged in a ring and interact via
a isotropic Heisenberg interaction $J_{i,j}^{\alpha }=J$.
The spins of the environment are all connected with each other and with all the spins of the system.
Previous work~\cite{YUAN06,YUAN07,YUAN08,YUAN09} has shown that it is expedient,
though not essential to take for the spin-spin interactions
$\Delta_{i,j}^{\alpha }$, and $\Omega_{i,j}^{\alpha }$
uniform random numbers in the range
$[-\left\vert \Delta\right\vert,\left\vert \Delta\right\vert ]$, and
$[-\left\vert \Omega\right\vert,\left\vert \Omega\right\vert ]$, respectively.
Relative to other choices of these interactions,
the randomness of the interaction parameters $\Delta$ and $\Omega$ and the high connectivity of the spins generally reduce the decoherence
and relaxation time to reach the stationary state of the reduced density matrix.
Note that we do not put any restriction on the energy spectra of the Hamiltonians describing $S$ and $E$.

\subsection{Initial state}

We prepare the state of the system $S$ and of the environment $E$ separately at $t<0$
and then bring them in contact with each other at $t=0$.
Specifically, we construct the initial pure state of the composite system $S+E$, $\rho(0)=|\Psi(0)\rangle \langle\Psi(0)|$
where
\begin{eqnarray}
|\Psi(0)\rangle =
|S\rangle\otimes\frac{ e^{-\beta H_E/2}|\Phi_E\rangle}{\langle\Phi_E|e^{-\beta H_E}|\Phi_E\rangle^{1/2}}
,
\label{inistate}
\end{eqnarray}
with $|\Phi_E\rangle=\sum_i c_i |\phi_i\rangle$ denoting the state of the environment with the coefficients
$c_i$ generated randomly according to the prescription given in Ref.~\cite{HAMS00} and
$\{|\phi_i\rangle\}$ being an orthonormal set of basis states which, in our simulation software,
are the usual direct products of the spin up and down states.
Numerically, the imaginary-time propagation by $e^{-\beta H_E/2}$ is performed by means of a
Chebyshev polynomial algorithm~\cite{TALE84,LEFO91,IITA97a,DOBR03}.
To prepare the environment in its ground state ($\beta=\infty$), we use the standard Lanczos method~\cite{GOLU96}.

It follows directly from Ref.~\cite{HAMS00} that for any observable $X_E(t=0)$ of the environment
\begin{eqnarray}
\langle \Psi(0)| X_E(t=0) |\Psi(0)\rangle &\approx&
\mathbf{Tr}\rho_E X_E(t=0)
,
\end{eqnarray}
the approximation improving as the inverse square root of the dimension of the Hilbert space of the environment
(see Appendix).
Therefore, we may consider the state $|\Psi(0)\rangle$ as ``typical'' in the sense that
if we measure observables of the environment, their expectation values agree with those
obtained from the canonical distribution of the environment at the inverse temperature $\beta$.
Note that in practice, it is often sufficient to consider only one random state $|\Phi_E\rangle$ (see Appendix).

The assumption of random phases in the initial state has been instrumental
in the derivation of the quantum master equation~\cite{HOVE55,HOVE57},
a key equation in the theory of non-equilibrium statistical mechanics.
Within the quantum master equation approach, the approach to equilbrium
of a quantum system is well understood~\cite{HOVE57,KUBO85}.
Although there may be an apparent similarity with the use of the random initial states
that we use in the present work, there is no relation between
the random initial states and the random phase assumption in the derivation of the master equation.
In the present work, random initial states are a convenient computational
device only: As we show in the Appendix, their use effectively eliminates the need
to compute traces of operators and allows us to work with pure states only.
Below, we also demonstrate explicitly that the use of random initial states is not essential
for the main conclusions of this paper by starting the simulation from the initial state with all spins up.

\subsection{Time evolution}

A pure state of the composite system $S+E$ evolves in time according to (in units of $\hbar=1$)
\begin{eqnarray}
|\Psi(t)\rangle&=&e^{-itH}|\Psi(0)\rangle=\sum_{i=1}^{D_s} \sum_{p=1}^{D_E} c(i,p,t)|i,p\rangle
,
\label{eq4}
\end{eqnarray}%
where the states $\{ |i,p\rangle \}$ are just another notation
of the complete set of orthonormal states in the spin-up -- spin-down basis
and $D_S=2^{n_s}$ and $D_E=2^{n_E}$ denote the dimension of the Hilbert space
of the system and environment, respectively.

Numerically, the real-time propagation by $e^{-itH}$ is carried out by means
of the Chebyshev polynomial algorithm~\cite{TALE84,LEFO91,IITA97a,DOBR03},
thereby solving the TDSE for the composite system starting from the initial state $|\Psi(0)\rangle$.
This algorithm yields results that are very accurate
(close to machine precision), independent of the time step used~\cite{RAED06}.

\subsection{Reduced density matrix}

The state of the quantum system $S$ is described by the reduced density matrix
\begin{equation}
\widetilde\rho(t)\equiv\mathbf{Tr}_{E}\rho \left( t\right)
,
\label{eq1}
\end{equation}
where $\rho \left( t\right) $ is the density matrix of the composite system at time $t$
and $\mathbf{Tr}_{E}$ denotes the trace over the degrees of freedom of the environment.
The system $S$ is in the canonical state if the reduced density matrix
takes the form
\begin{equation}
\widehat\rho (\beta )\equiv\left.{e^{-\beta H_{S}}}\right/{\mathbf{Tr}_{S}e^{-\beta H_{S}}}
,
\label{eq2}
\end{equation}
where $\mathbf{Tr}_{S}$ denotes the trace over the degrees of freedom of the system $S$.
In terms of the expansion coefficients $c(i,p,t)$, the matrix element $(i,j)$ of the reduced density matrix reads
\begin{eqnarray}
\widetilde\rho_{i,j}(t) &=&\mathbf{Tr}_{E} \sum_{p=1}^{D_E}\sum_{q=1}^{D_E} c^\ast(i,q,t)c(j,p,t)|j,p\rangle\langle i,q|
\nonumber \\
&=&\sum_{p=1}^{D_E} c^\ast(i,p,t)c(j,p,t)
.
\end{eqnarray}

\subsection{Data analysis}

We analyze the time-dependent data of the reduced density matrix in various ways.
First, at each time step (in units of $\tau=\pi/10$),
we diagonalize the (non-negative definite) reduced density matrix itself and study
the time-dependence of its eigenvalues.
We define the variance of the set of eigenvalues at $t$ and $t_f$ by
\begin{equation}
\mathrm{var}(t)\equiv \sqrt{ \sum_{i=1}^{D_S} (\lambda_i(t)- \lambda_i(t_f))^2},
\label{eqvar}
\end{equation}
where $\lambda_i(t)$ is the $i$th eigenvalue of $\widetilde{\rho}(t)$.
Usually, $t_f$ is taken to be the final time of the simulation.
From the eigenvalues, we also compute the entropy of the system
\begin{equation}
S(t)\equiv -\mathbf{Tr} \widetilde{\rho}(t) \ln \widetilde{\rho}(t) = -\sum_{i=1}^{D_S} \lambda_i(t) \ln \lambda_i(t).
\label{eqentropy}
\end{equation}

We characterize the degree of decoherence of the system by
\begin{equation}
\sigma (t) =\sqrt{\sum_{i=1}^{D_S-1}\sum_{j=i+1}^{D_S}\left\vert\widetilde\rho_{ij}(t) \right\vert ^{2}},
\label{eqsigma}
\end{equation}%
where $\widetilde\rho_{ij}(t)$ is the matrix element $(i,j)$
of the reduced density matrix $\widetilde\rho$ in the representation
that diagonalizes $H_S$.
Clearly, $\sigma(t)$ is a global measure for the size of the
off-diagonal terms of the reduced density matrix in the representation
that diagonalizes $H_S$.
If $\sigma(t)=0$ the system is in a state of full decoherence
(relative to the representation that diagonalizes $H_S$).

\begin{figure}[t]
\begin{center}
\includegraphics[width=7cm]{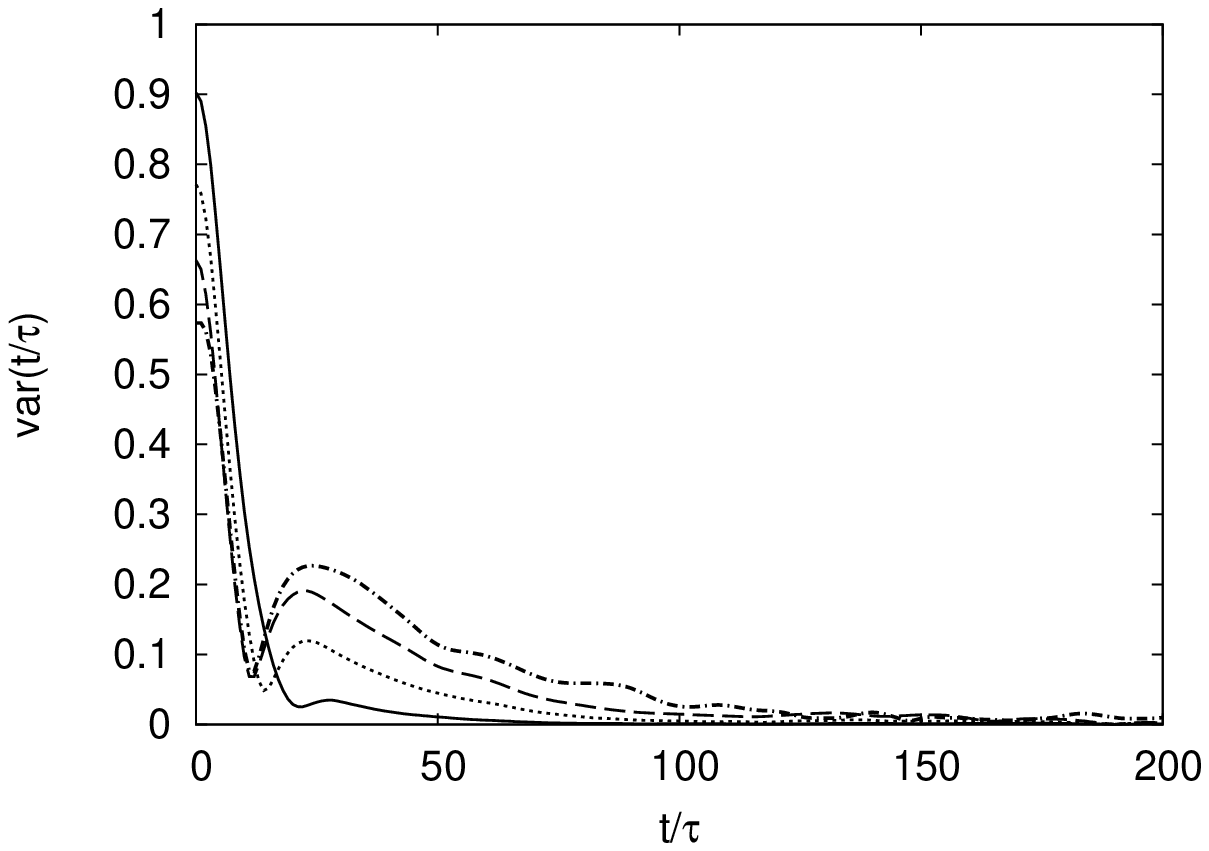}
\includegraphics[width=7cm]{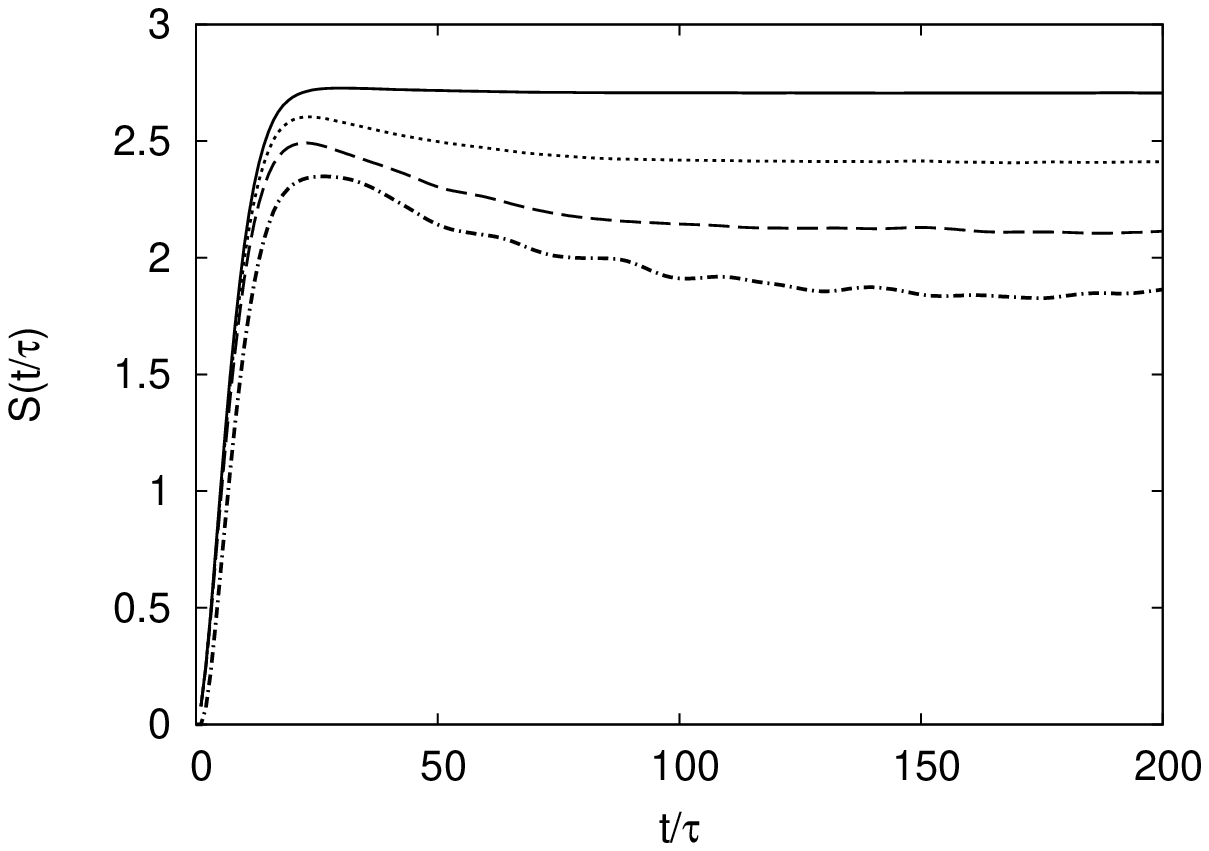}
\includegraphics[width=7cm]{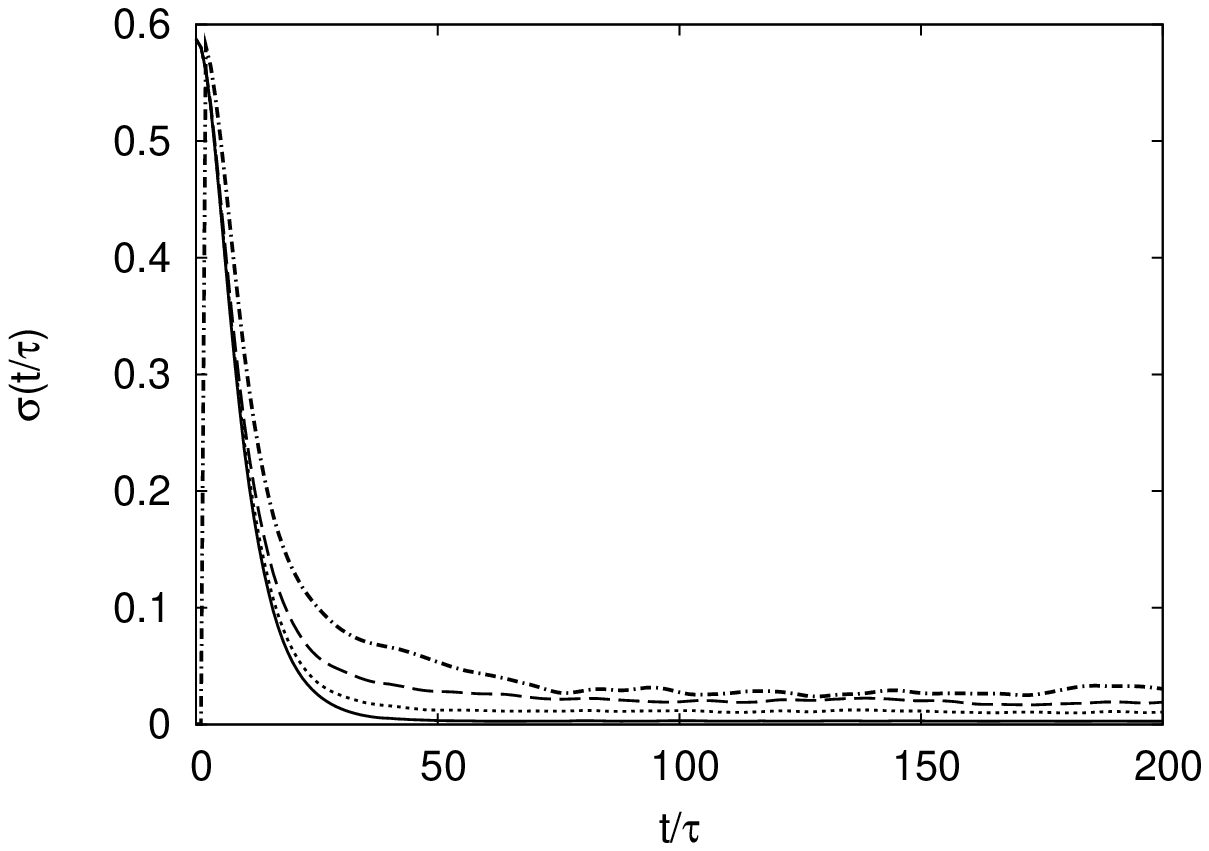}
\includegraphics[width=7cm]{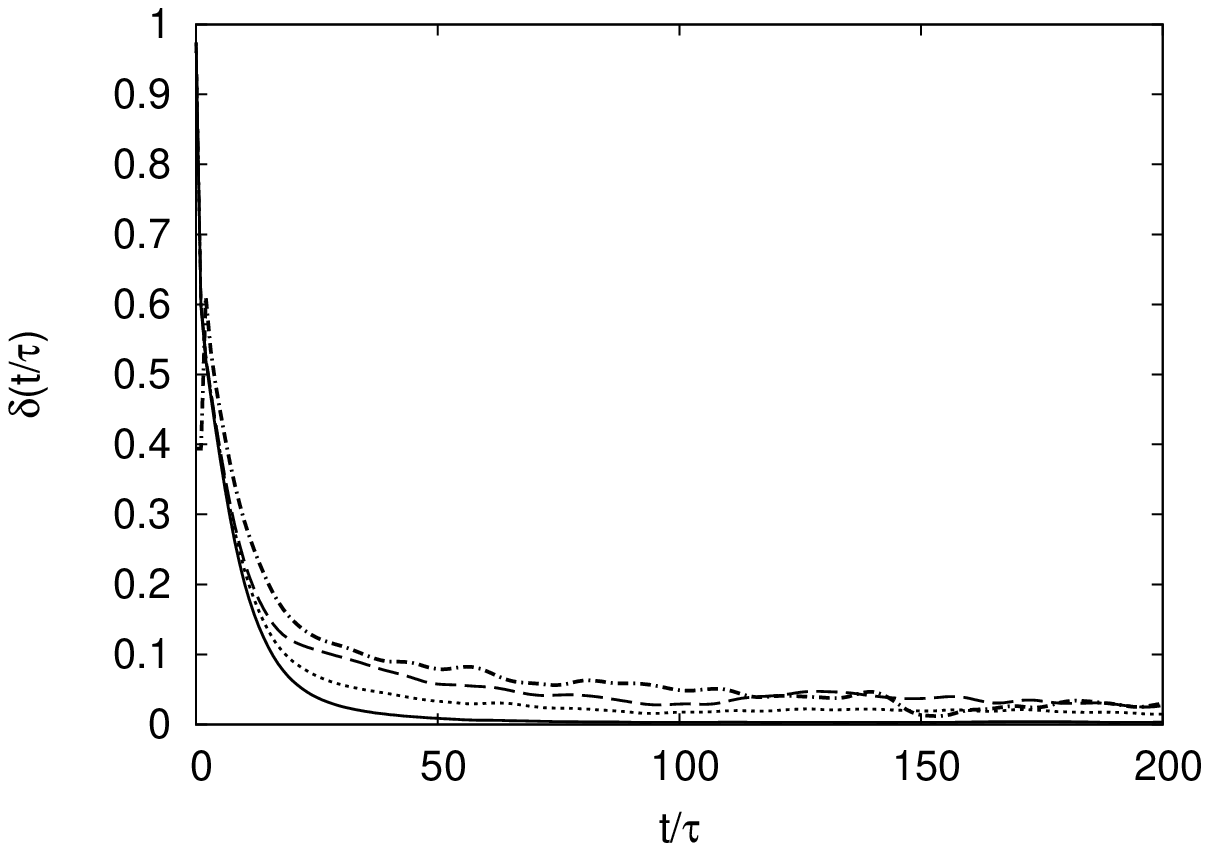}
\caption{%
Simulation results for $\mathrm{var}(t/\tau)$ and $S(t/\tau)$
(see Eqs.~(\ref{eqvar}) and (\ref{eqentropy}))
and $\sigma(t/\tau)$ and $\delta(t/\tau)$ (see Eqs.~(\ref{eqsigma}) and (\ref{eqdelta}))
for different values of the initial temperature of the environment,
as obtained by solving the TDSE for a system consisting of a ring of four spins
coupled to an environment of eighteen spins.
The spins of the environment are all connected with each other and with the four spins of the system.
The initial state of the system $S$ is given by $|\uparrow\downarrow\uparrow\downarrow\rangle$ and the
initial state of the environment $E$ is given by a random pure state.
Model parameters: $J=-0.5$, $\Delta=0.2$, $\Omega=0.5$, $t_f/\tau=190$;
Sample interval $\tau=\pi/10$.
Solid line: $\beta=1$;
Dotted line: $\beta=3$;
Dashed line: $\beta=6$;
Dash-dotted line: $\beta=\infty$.
}
\label{sfig2}
\end{center}
\end{figure}

Assuming that the system $S$, evolving
in time according to the TDSE, relaxes to the canonical state
we expect that $\widetilde\rho\left( t\right)\approx\widehat\rho (b)$
for $t>t_0$ where $t_0$ is some finite time and $b$ denotes the effective inverse temperature of $S$.
The difference between the state $\widetilde\rho\left( t\right)$
and the canonical distribution $\widehat\rho (b(t))$
is conveniently characterized by
\begin{equation}
\delta(t)=\sqrt{\sum_{i=1}^{D_S}\left( \widetilde\rho_{ii}(t) -
\left.{e^{-b(t) E_{i}}}\right/{\sum_{i=1}^{D_S} e^{-b \left( t\right) E_{i}}}\right) ^{2}}
,
\label{eqdelta}
\end{equation}%
with
\begin{equation}
b(t)=\frac{\sum_{i<j,E_{i}\neq E_{j}}
[\ln \widetilde\rho_{ii}(t) -\ln \widetilde\rho _{jj}(t)]/({E_{j}-E_{i}})}{\sum_{i<j,E_{i}\neq E_{j}}1}.
\label{eqbt}
\end{equation}%
If the system relaxes to its canonical distribution
both $\delta(t)$ and $\sigma (t)$ are expected to vanish,
$b(t)$ converging to the effective inverse temperature $b$.

For any function $f(.)$ of the system Hamiltonian $H_S$,
we define the averages
\begin{eqnarray}
\langle f(H_S) \rangle_{{\widetilde\rho}(t)}& \equiv& \mathbf{Tr}\widetilde\rho(t)f(H_S)
,
\end{eqnarray}
and
\begin{eqnarray}
\langle f(H_S) \rangle_{b}& \equiv& \left. \mathbf{Tr}e^{-bH_S}f(H_S)\right/\mathbf{Tr }e^{-bH_S}
.
\label{fH}
\end{eqnarray}
Then, application of the Schwarz inequality yields
\begin{eqnarray}
\left| \langle f(H_S) \rangle_{\widetilde\rho(t)}
-
\langle f(H_S) \rangle_{b} \right|^2 &\le& \delta^2(t) \mathbf{Tr}f^2(H_S)
,
\end{eqnarray}
showing that the deviations of the energy and entropy of the system from
their values in the canonical ensemble vanish linearly or faster with $\delta(t)$.

\section{Results}
Most of our simulations have been carried out for systems consisting of four spins coupled to an environment
of eighteen spins.
We have verified that our conclusions do not depend on details such as the connectivity of the spins
in the environment or the size of the composite system by simulating
triangular lattices, regular square lattices and so on with up to 35 spins (data not shown).

\begin{table}[t]
\caption{%
Data for $\sigma$, $\delta$, $S_{\tilde{\rho}}$ and $E_{\tilde{\rho}}$,
taken at the last time step of the simulation run.
$S_b$ and $E_b$ are calculated according to Eq.~(\ref{fH}).
}
\label{tab00}       
\begin{tabular}{lccccccc}
\hline
\noalign{\smallskip}
$\beta$ & $b$ &$\sigma$ & $\delta$ & $S_b$ & $S_{\tilde{\rho}}$ & $E_b$ & $E_{\tilde{\rho}}$\\
\hline
&\multicolumn{6}{c}{$J=-0.5$, $\Delta=0.2$, $\Omega=0.5$}  &\\
\hline
1 &  0.807 & 0.003  & 0.003  & 2.704 & 2.706 & -0.165 & -0.162\\
3	& 1.851	&	0.011	&0.015	 &	2.392	&	2.412	 &-0.400	&	-0.388\\
6	&2.526	&	0.019	&0.027	&	2.080	&	2.114	&-0.543 &	-0.526\\
$\infty$ &	3.044	&	0.031	&0.030	&	1.816 &	1.852 &	-0.638 &-0.621\\
\noalign{\smallskip}
\hline
\end{tabular}
\end{table}

\begin{figure}[t]
\begin{center}
\includegraphics[width=7cm]{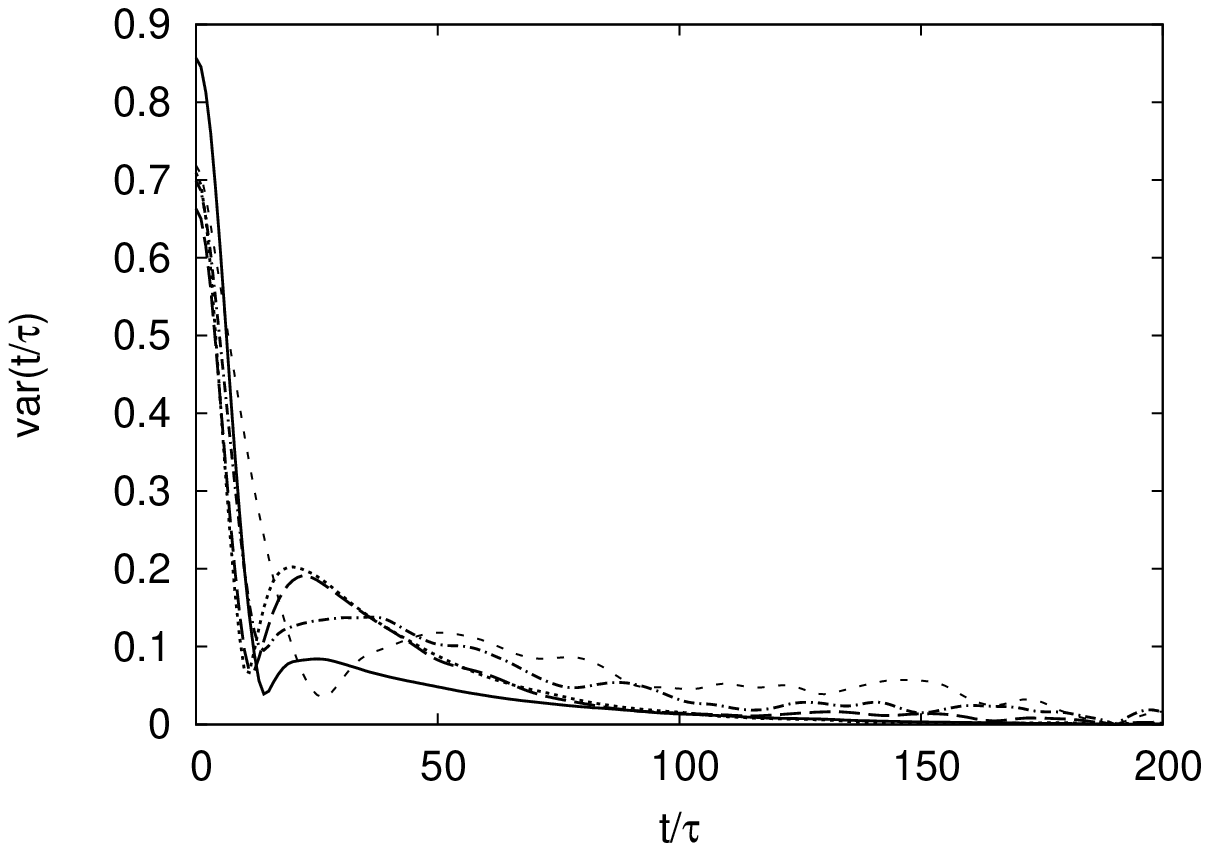}
\includegraphics[width=7cm]{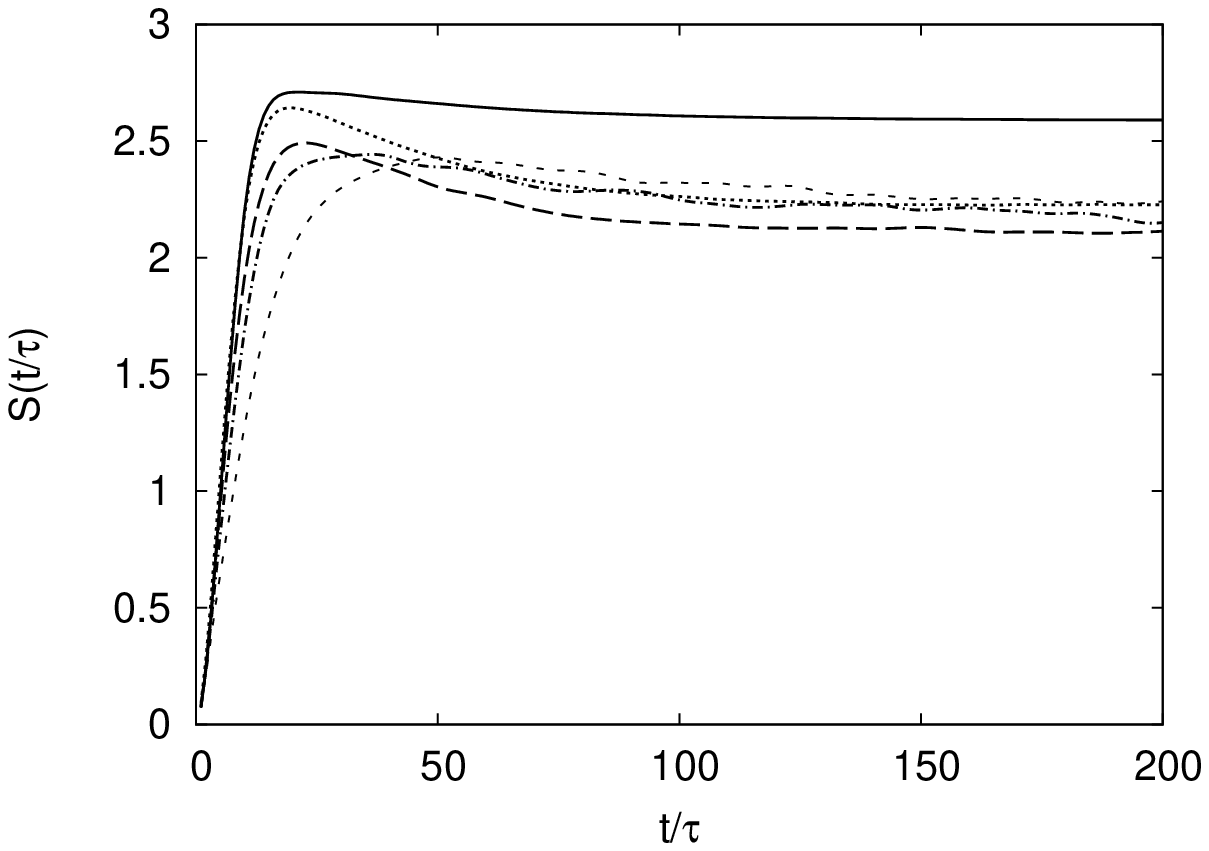}
\includegraphics[width=7cm]{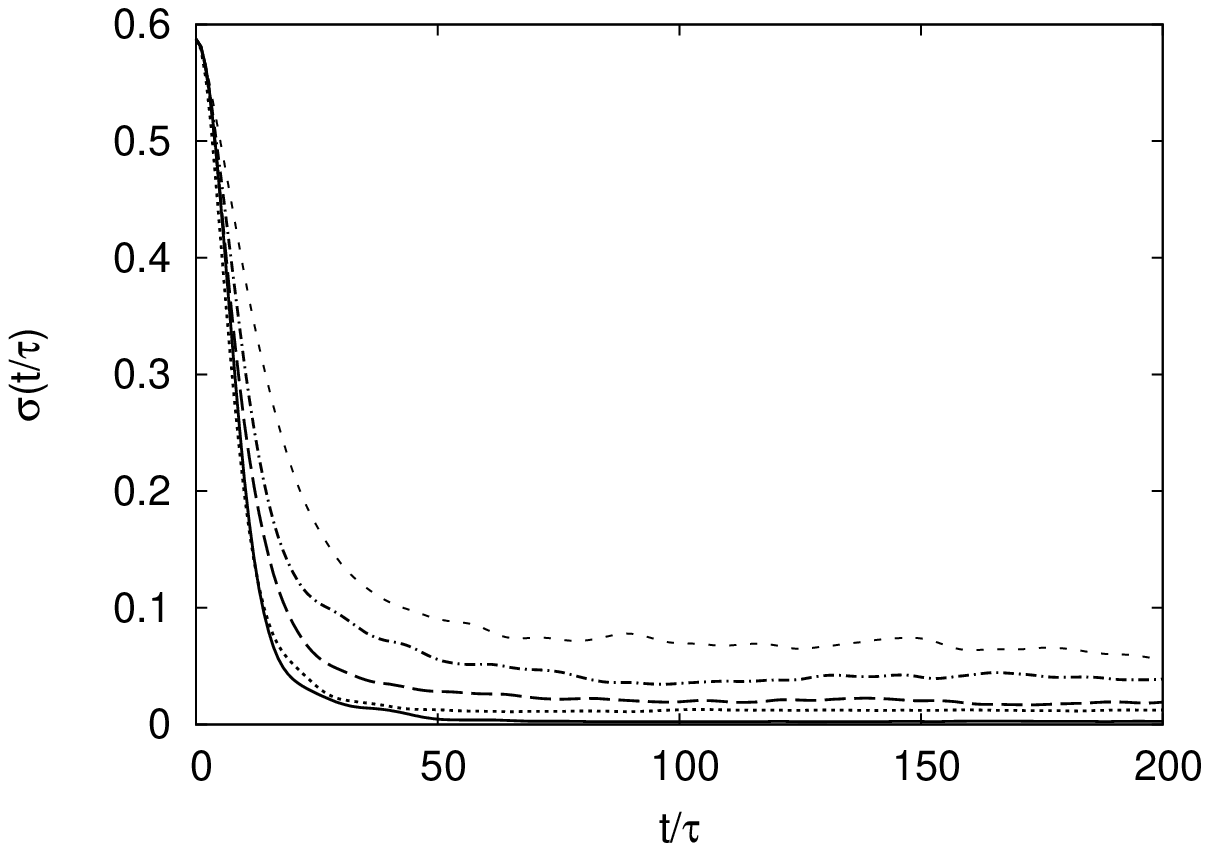}
\includegraphics[width=7cm]{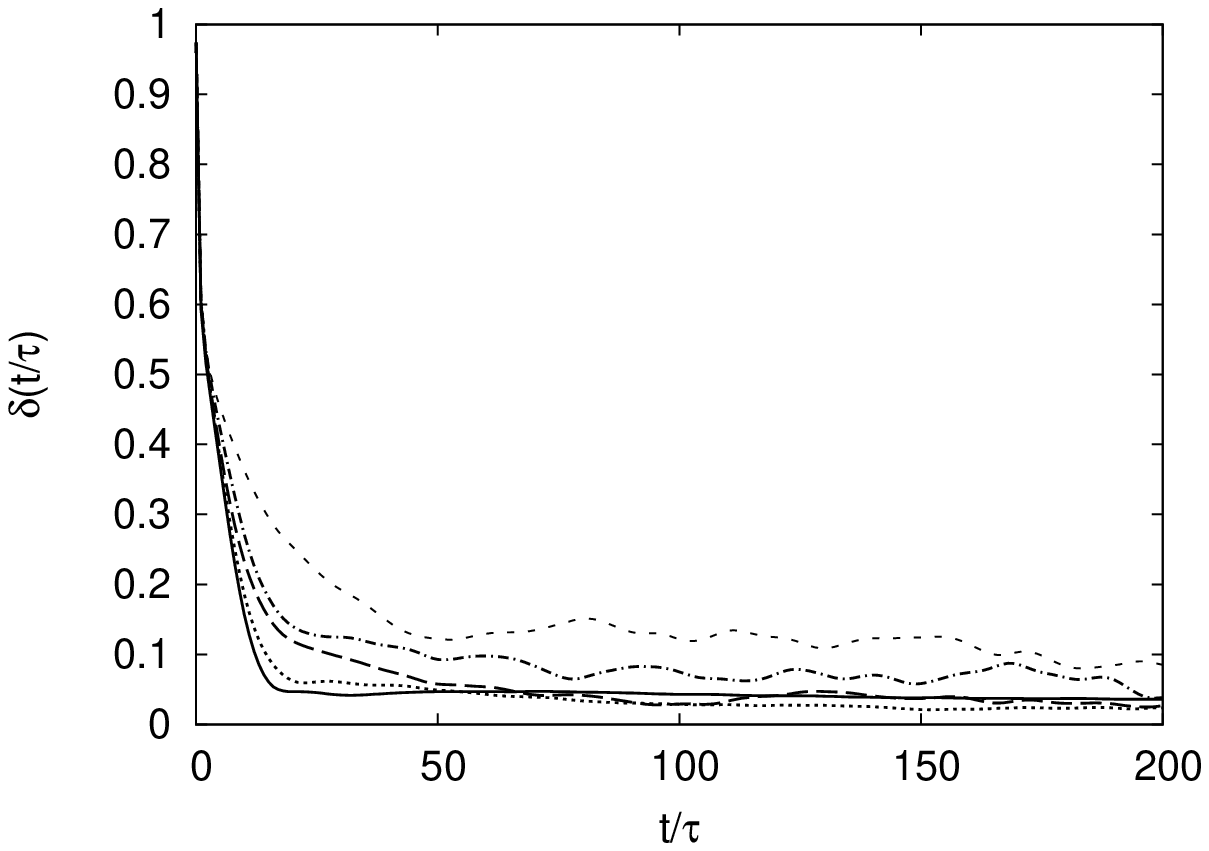}
\caption{%
Simulation results for $\mathrm{var}(t/\tau)$, $S(t/\tau)$, $\sigma(t/\tau)$, and $\delta(t/\tau)$
for different values of the energy scale of the couplings between the spins in the environment.
The data are obtained by solving the TDSE for a system consisting of a ring of four spins
coupled to an environment of eighteen spins.
The spins of the environment are all connected with each other and with the four spins of the system.
The initial state of the system $S$ is given by $|\uparrow\downarrow\uparrow\downarrow\rangle$ and the
initial state of the environment $E$ is given by a random pure state.
Model parameters: $J=-0.5$, $\Delta=0.2$, $\beta=6$, $t_f/\tau=190$;
Sample interval $\tau=\pi/10$.
Solid line: $\Omega=0.125$;
Dotted line: $\Omega=0.25$;
Long-dashed line: $\Omega=0.5$;
Dash-dotted line: $\Omega=0.75$.
Short-dashed line: $\Omega=1$.
}
\label{sfig3}
\end{center}
\end{figure}

\begin{figure}[t]
\begin{center}
\includegraphics[width=7cm]{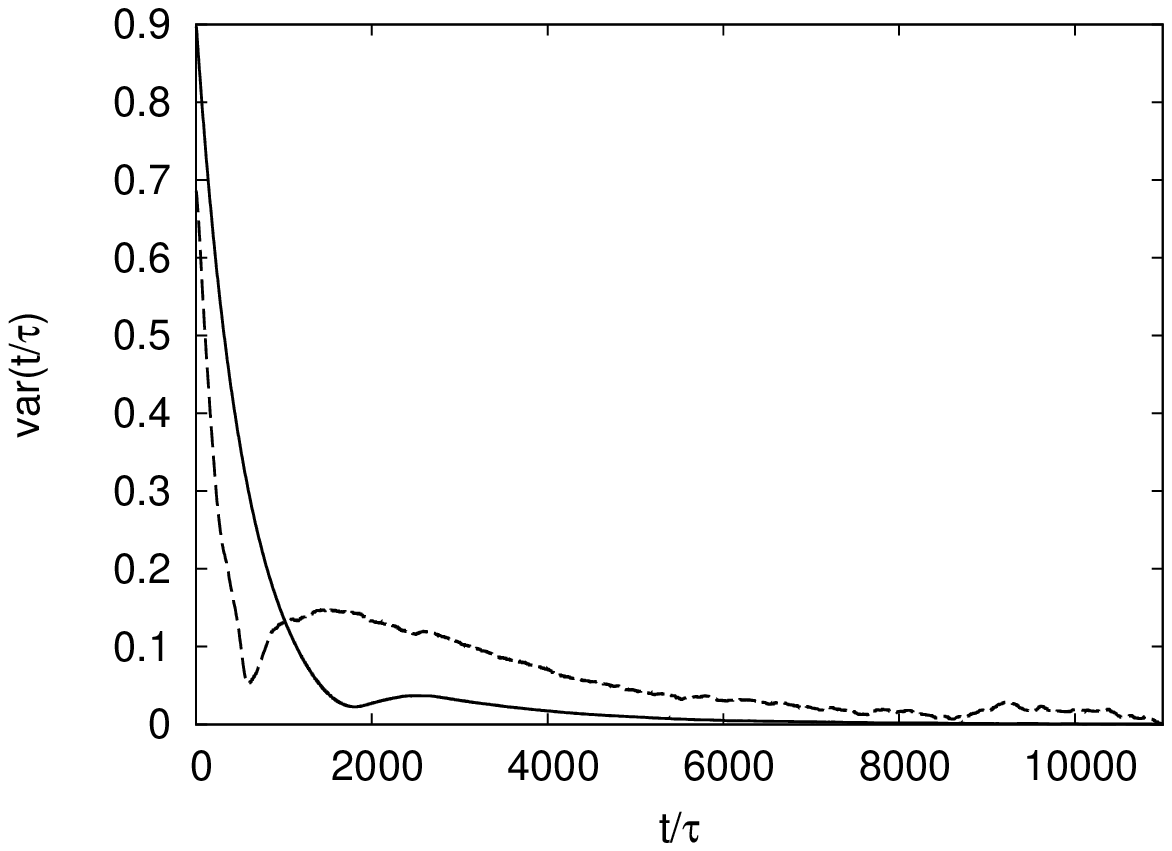}
\includegraphics[width=7cm]{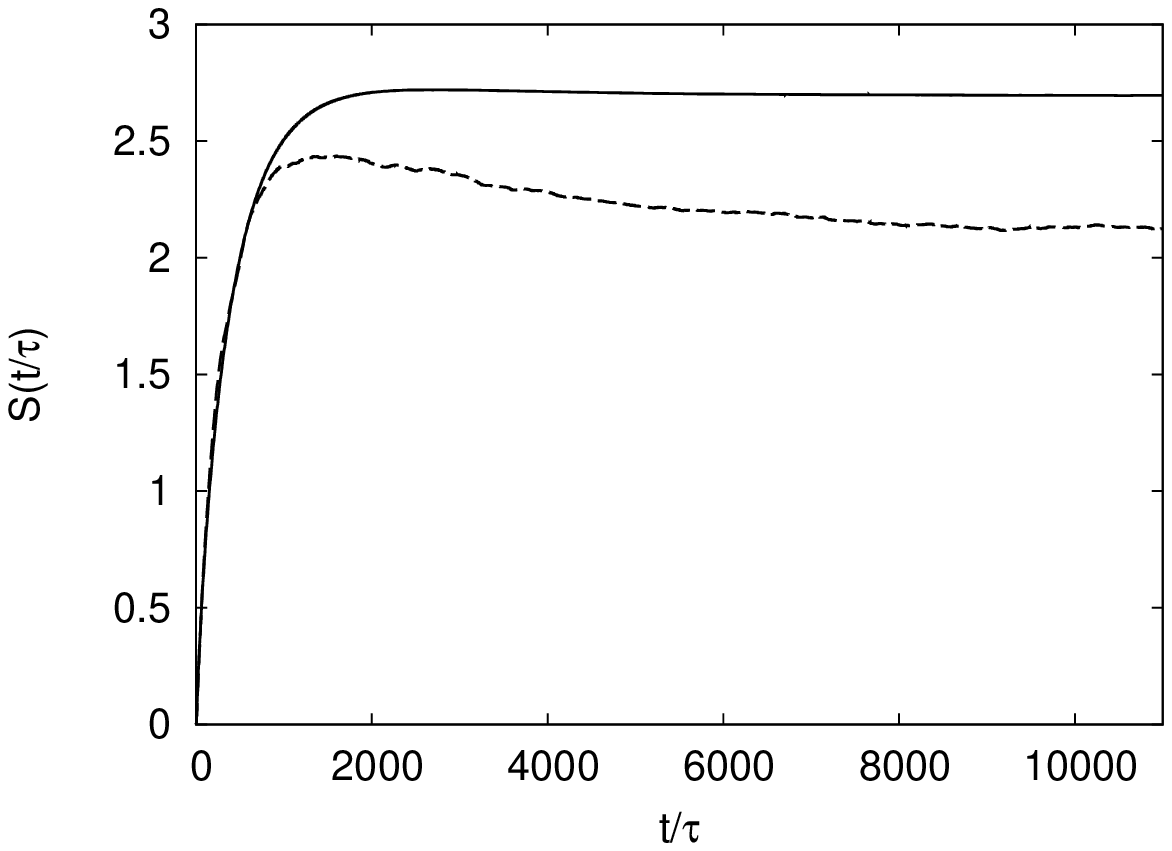}
\includegraphics[width=7cm]{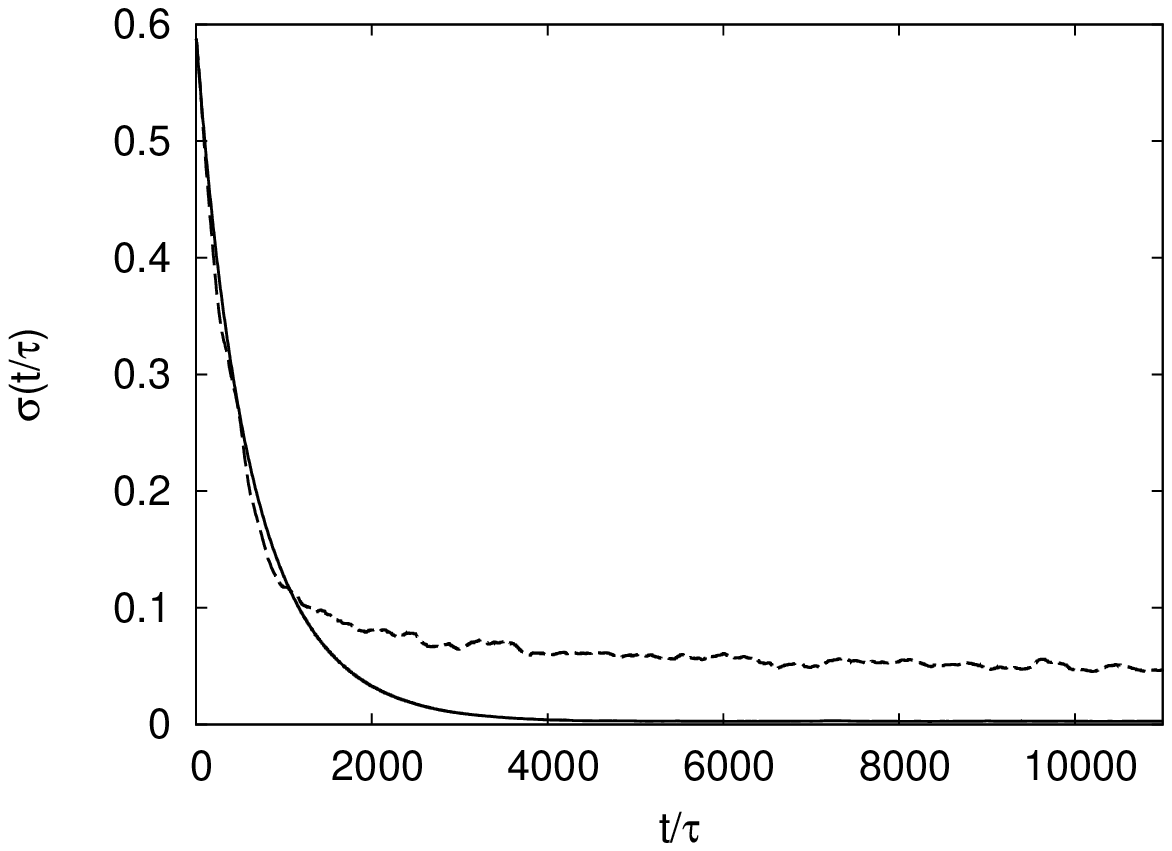}
\includegraphics[width=7cm]{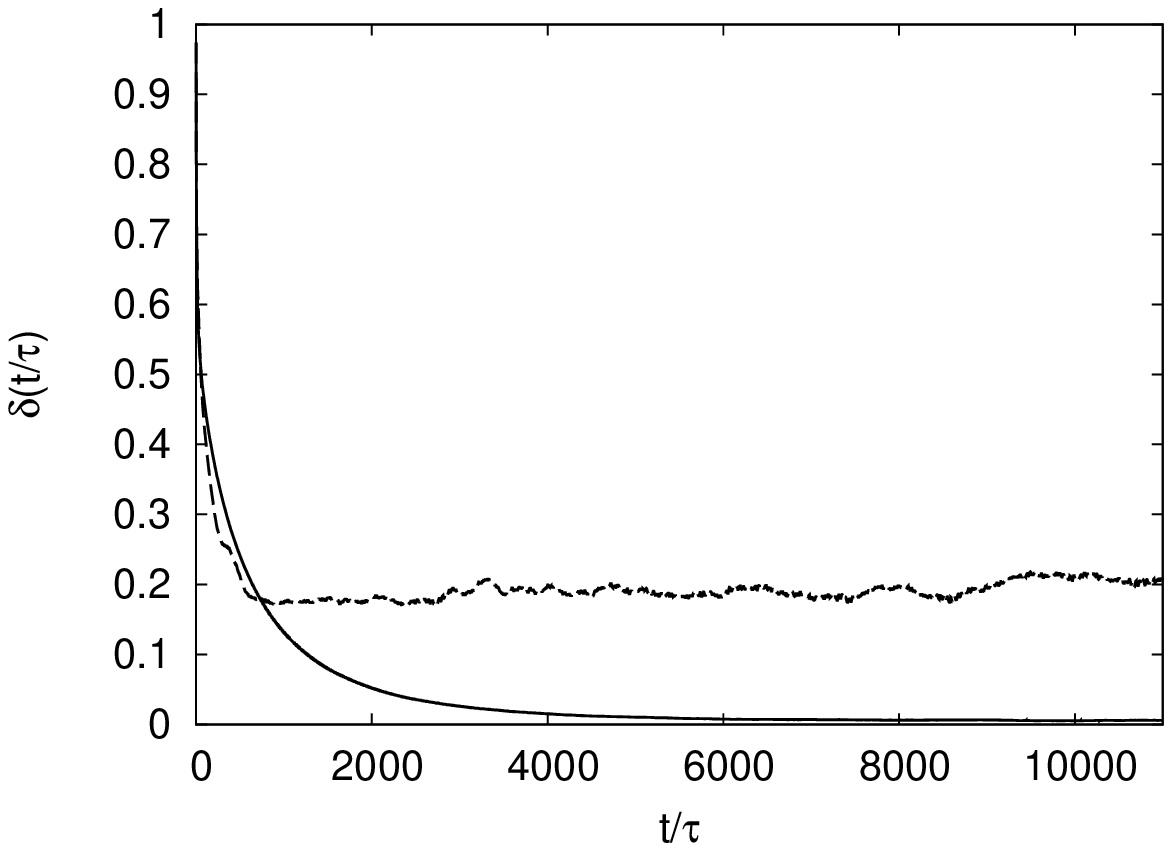}
\caption{%
Same as Fig.~\ref{sfig2} except that the
system-environment interaction strength $\Delta=0.02$.
Model parameters: $J=-0.5$, $\Delta=0.02$, $\Omega=0.5$, $t_f/\tau=11000$;
Sample interval $\tau=\pi/10$.
Solid line: $\beta=1$;
Dashed line: $\beta=6$.
}
\label{sfig4}
\end{center}
\end{figure}

In Fig.~\ref{sfig2}, we present the simulation results
for $\mathrm{var}(t/\tau)$, $S(t/\tau)$
(see Eqs.~(\ref{eqvar}) and (\ref{eqentropy})),
$\sigma(t/\tau)$ and $\delta(t/\tau)$ (see Eqs.~(\ref{eqsigma}) and (\ref{eqdelta}))
for the case that there is a fairly strong coupling between the system $S$ and the environment $E$ ($|\Delta/J|=0.4$)
and for different values of the initial temperature of the environment.
From Fig.~\ref{sfig2}, it is clear that, independent of the initial temperature of the environment,
all the eigenvalues $\lambda_i(t)$ of the
reduced density matrix converge to a stationary value.
This implies that also the entropy of the system $S$ approaches a stationary value,
suggesting that the system $S$ relaxes to an equilibrium state.
We emphasize that the data presented in this paper are obtained
without any averaging procedure other than the one intrinsic to quantum theory.

From Fig.~\ref{sfig2}, it follows from the data for $\sigma(t)$ that if $\beta=1$,
the reduced density matrix of the system $S$ converges to a diagonal matrix
in the representation that diagonalizes $H_S$,
in other words, the system $S$ has lost virtually all coherence ($\sigma\rightarrow 0$)
and as $\delta\rightarrow 0$ with time, the system $S$ relaxes to the canonical system.
The same holds for $\beta =0$, see Ref.~\cite{YUAN09}.
With increasing $\beta$, the difference between the reduced density matrix and the canonical distribution
(for the system defined by $H_S$) increases slightly,
as indicated by the fact that the values of $\sigma$ and $\delta$ increase with $\beta$.

It is instructive to analyze more quantitatively, the data taken at the end of these particular
simulation runs. In Table~\ref{tab00} we collect the results for the various quantities
of interest.
As $\beta$ increases, $\sigma$ increases too, indicating that the deviation from a diagonal matrix
(with respect to the basis that diagonalizes $H_S$) increases, merely reflecting
the fact that the decoherence processes become less effective as the temperature decreases.
Nevertheless, even at zero temperature, the difference between the reduced density matrix
and the canonical density matrix is quite small, of the order af a few percent, and so are
the differences for the entropy and energy.
Thus, it seems that even for fairly strong coupling between the system and the environment
($|\Delta/J|=0.4$), the nanoscale environment drives the even smaller system to a state
that, within a few percent, is described by the canonical state of the system, albeit with
an effective temperature that does not agree with that of the environment (compare $\beta$
and $b$ in Table~\ref{tab00}).

\subsection{Energy scale of the environment}\label{subsec}
To investigate the effect of the parameter $\Omega$, the energy scale of the states of the environment,
we performed simulations for $\Omega=0.125,0.25,0.5,0.75,1$.
In Fig.~\ref{sfig3} we present data for $\beta=6$.
From earlier work~\cite{YUAN06,YUAN07,YUAN08} for $\beta=0$ and $\beta=\infty$ we know
that the more the range and the width of the energy spectrum of the environment and the system match each other,
the better the system decoheres. Taking this into account we can easily understand the behavior of $\sigma$:
When the value of $\Omega$ decreases, the value of $\sigma$ decreases.

Except for $\Omega<|J|$ (recall $J=-0.5$ in this paper), both $\sigma(t)$ and $\delta(t)$ relax to fairy small values,
the difference between the reduced density matrix at $t/\tau=200$ with the canonical states being of the order
of a few percent (data not shown).
For $\Omega>|J|$, the qualitative behavior is different: Although $S(t)$, $\sigma(t)$ and  $\delta(t)$
to relax to their stationary values, the difference between the reduced density matrix and the canonical state
is significant, as indicated by $\delta(t/\tau=200)\approx0.1$.
Qualitatively, this can be understood as follows. Keeping the number of spins of the environment
and the range of $S-E$ interactions $\Delta$ fixed,
increasing $\Omega$ increases the spectral range of the environment,
that is the spacing between the energy levels of the environment increases.
This effectively reduces the mixing of the eigenstates of the system and the environment
by the $S-E$ interactions, which in turn leads to a reduction of the effect of decoherence by the environment
and the probabilities for the system to exchange energy with the environment.

\begin{table}[t]
\caption{%
Same as Table~\ref{tab00} except that the coupling $\Delta=0.02$ instead of $\Delta=0.2$.
}
\label{tab02}       
\begin{tabular}{lccccccc}
\hline
\noalign{\smallskip}
$\beta$ & $b$ &$\sigma$ & $\delta$ & $S_b$ & $S_{\tilde{\rho}}$ & $E_b$ & $E_{\tilde{\rho}}$\\
\hline
&\multicolumn{6}{c}{$J=-0.5$, $\Delta=0.02$, $\Omega=0.5$}  &\\
\hline
1	& 0.874	&	0.003 &	0.005	&	2.692	&	2.696	&-0.180	&	-0.175 \\
6	& 3.349 &	0.047	& 0.207	&	1.659	&	2.125	&-0.687	&	-0.495\\
\noalign{\smallskip}
\hline
\end{tabular}
\end{table}

\subsection{Weak interaction between system and environment}
For the simulation data presented earlier, the system-environment interaction strength $\Delta$
was chosen such that $|\Delta/J|=|\Delta/\Omega|={\cal O}(1)$ which is far away from the weak coupling regime.
If we reduce $\Delta$, we may expect that the time scale of decoherence and relaxation processes
increases with $\Delta$.
To keep the amount of computer time required to reach the stationary state within reasonable limits,
we have chosen to reduce $\Delta$ by a factor ten, that is we take $\Delta=0.02$
and consider this to be the ``weak coupling'' case.
The simulation results for this parameter choice are presented in Fig.~\ref{sfig4}.
Note that compared to Fig.~\ref{sfig2} and Fig.~\ref{sfig3}, the time scale to reach the stationary state
has increased by a factor of about one hundred.
For $\beta=1$, there is no qualitative change compared to the case of $\Delta=0.2$:
The state of the system converges to the canonical state.

However, for $\beta=6$ the relatively large values of $\sigma(t)$ and $\delta(t)$ signal that
the decoherence process is not very effective and that the deviation from the canonical distribution
is significant, as is shown quantitatively in Table~\ref{tab02}.
Nevertheless, the eigenvalues of the reduced density matrix relax to their stationary values and so does
the system entropy.
Qualitatively, this can be understood by the same argument as the one used at the end of Section~\ref{subsec}.
Keeping the number of spins of the environment
and the spectral range of the environment $\Omega$ fixed,
decreasing the range of $S-E$ interactions $\Delta$ effectively reduces
the mixing of the eigenstates of the system and the environment,
which in turn leads to a reduction of the effect of decoherence by the environment
and the probabilities for the system to exchange energy with the environment.

\begin{figure}[t]
\begin{center}
\includegraphics[width=7cm]{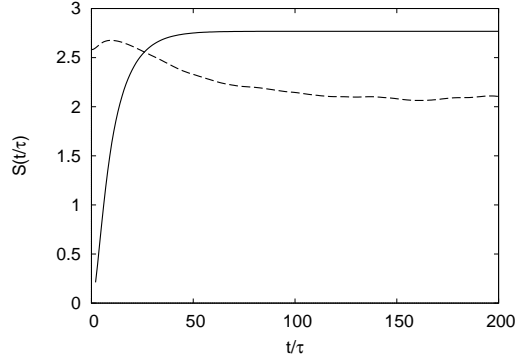}
\caption{%
The entropy of the system as a function of $t/\tau$ for the same system as in Fig.~\ref{sfig2}
except that the initial state of the system and environment is the state with all spins up (solid line)
or  $|\Psi(0)\rangle =e^{-\beta H_E/2}|\Phi\rangle/\langle\Phi|e^{-\beta H_E}|\Phi\rangle^{1/2}$ (dashed line)
where $|\Phi\rangle$ is a uniform random superposition of all states of the whole system and $\beta=6$.
}
\label{sfig5a}
\end{center}
\end{figure}

\subsection{Properties of the stationary state}

The observation that the eigenvalues of the reduced density matrix,
and therefore also the entropy of the system,
approach stationary values for sufficiently
long times seems to be generic, independent of the values of
the parameters $\Delta/|J|\le1$, $\Omega/|J|\le1$, $\beta$ and the initial state of the system $S$
or the environment $E$, as illustrated in Fig.~\ref{sfig5a}.
As a matter of fact, in our large collection of simulation results (most results not shown)
there is no evidence of the contrary.
However, the stationary state itself depends on all the above mentioned parameters.
The time scale on which the equilibration occurs strongly depends on the system-environment interaction strength $\Delta$.
As expected, reducing $\Delta$ increases the time scale of decoherence and relaxation processes.
Although our simulation results are obtained for a very small quantum system connected to a relatively small environment, they
are in good agreement with findings for interacting large quantum systems that seem to evolve in such a way that
any small subsystem equilibrates, under the condition that the Hamiltonian has no degenerate energy gaps and that the state of the
composite system contains sufficiently many eigenstates~\cite{LIND09}.

\begin{figure}[t]
\begin{center}
\includegraphics[width=7cm]{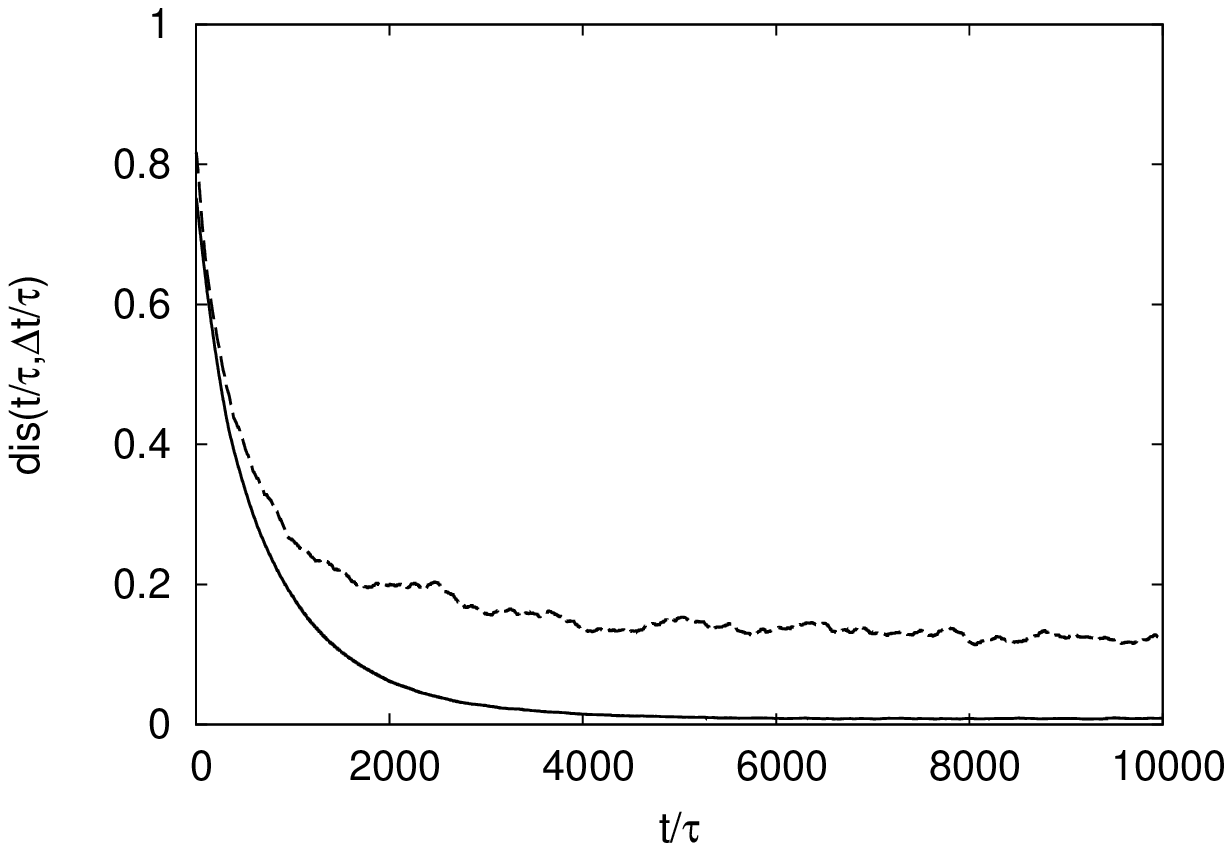}
\includegraphics[width=7cm]{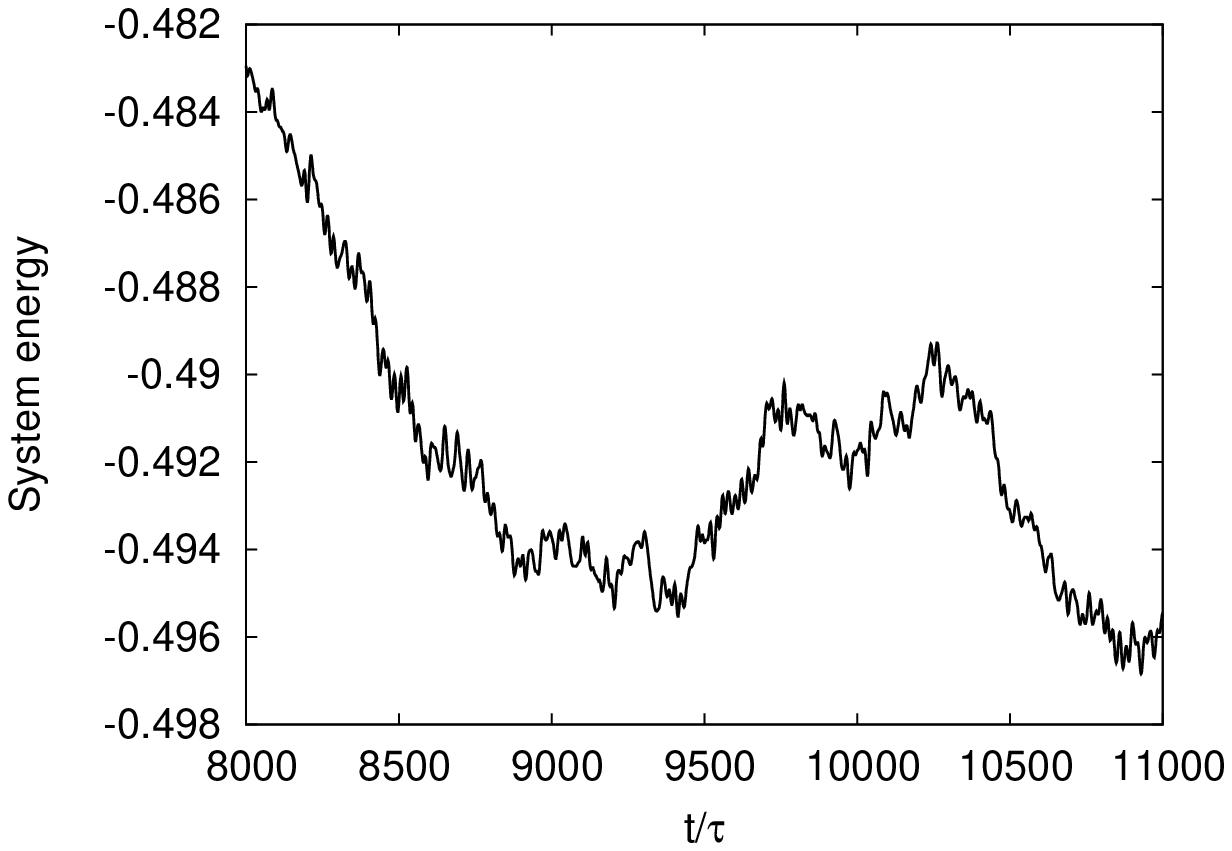}
\caption{%
Left:The difference $\mathrm{dis}(t/\tau,\Delta t/\tau)=\mathbf{Tr} |\widetilde\rho(t)-\widetilde\rho(t+\Delta t)|/2$
between the reduced density matrices
$\widetilde\rho(t)$ and $\widetilde\rho(t+\Delta )t$ as a function of $t/\tau$
for a fixed value $\Delta t/\tau=1000$,
extracted from the simulations that yield the data presented in Fig.~\ref{sfig4}.
Solid line: $\beta=1$;
Dashed line: $\beta=6$.
Right: The energy of the system as a function of $t/\tau$ for a time interval chosen
in which there is exchange of energy between the system and the environment,
showing that the dynamical evolution of the system is non-Markovian.
}
\label{sfig5}
\end{center}
\end{figure}

However, our simulation results also show that the eigenstates of the reduced density matrix
generally do not evolve to a stationary state.
In Fig.~\ref{sfig5}(left) we show representative results of a measure for the distance,
the trace distance
$\mathrm{dis}(t/\tau,\Delta t/\tau)=\mathbf{Tr} |\widetilde\rho(t)-\widetilde\rho(t+\Delta t|/2$,
between the two reduced density matrices $\widetilde\rho(t)$ and $\widetilde\rho(t+\Delta t)$ as a function of $t$
for a fixed value of $\Delta t$, as obtained from the simulations that yield the data
presented in Fig.~\ref{sfig4}.
For $\beta=1$, the case in which the difference between the reduced density matrix
and the canonical state is small, $\mathrm{dis}(t/\tau,\Delta t/\tau)$
becomes very small for $t/\tau>8000$, suggesting that
the eigenvectors of the reduced density matrix also converge to
their stationary values (in concert with the fact that in this case,
the reduced density matrix is very close to the canonical state).
In contrast, for $\beta=6$ and $t/\tau$ large, $\mathrm{dis}(t/\tau,\Delta t/\tau)$
levels off at a value which is not small.
In other words, although the eigenvalues of the reduced density matrix
relax to their stationary values, the eigenvectors of the reduced density matrix
exhibit nontrivial quantum dynamics, even though the entropy of the system
has reached a stationary, nonzero, value.

In the case of a single spin and two-spins interacting with an environment,
the dynamics of eigenvectors of the reduced density matrix in the stationary-state
was observed earlier through quantum oscillations of spin expectation values~\cite{DOBR03z,MELI04}.
In Fig.~\ref{sfig5}(right), we present results for the energy of the subsystem, as obtained
from the simulation that produced the data of Fig.~\ref{sfig4}(right).
It is clear that the system energy mostly decreases, with energy going from the system to the environment.
However, for some time intervals (one being shown), the system energy increase, indicating
that the environment transfers energy to the system,
a characteristic feature of non-Markovian processes~\cite{BREU09,BREU10}.

\section{Conclusion}%
We have presented a simulation study of a small magnetic system coupled to a nanoscale magnetic environment.
The quantum dynamical evolution of the composite system was obtained from the direct numerical solution of the TDSE.
Our analysis, albeit numerical, does not involve any approximation,
does not rely on time-averaging of observables
nor does it assume that the coupling between system and environment is weak.

The most striking result of our analysis is that
all the eigenvalues of the reduced density matrix relax to stationary values,
implying that the entropy of the system relaxes to a stationary value.
In this sense, the nanoscale environment drives the system to a thermodynamically stationary state,
a feature which is usually attributed to macroscopic environments only.

Furthermore, we have shown that under suitable but fairly general conditions, the reduced density matrix in the stationary
state is close to the canonical state of the system,
albeit not with the same temperature as the one of the environment.
The difference between these two states generally increases as the temperature of the environment decreases.

For $\beta\approx0$, the initial state of the environment has all the features of a ``canonical typicality'' state
and qualitatively, our results are in concert with the theory of canonical typicality:
The reduced density matrix relaxes to the canonical state~\cite{GOLD06,POPE06}.
As $\beta$ increases the concept of ``canonical typicality'' no longer applies in the strict sense,
as explained in the Appendix.
Yet, qualitatively, we may interpret our findings in terms of the effective dimension $d_E^{eff}$
of the environment~\cite{POPE06}:
If $\beta\approx0$,  $d_E^{eff}={\cal O}(D_E)$, which is a large number for the systems that we have used
in our simulations. As $\beta$ increases, the number of states of the environment effectively available
for decoherence and relaxation
decreases (see also the results of the density of states in Fig.~\ref{sfig6}).
As this implies that $d_E^{eff}$ decreases, it is to be expected that the difference between
the reduced density matrix and the canonical state increases~\cite{POPE06}.

\section*{Acknowledgement}
This work is partially supported by NCF, the Netherlands,
by a Grant-in-Aid for Scientific Research on Priority Areas,
and the Next Generation Super Computer Project, Nanoscience Program from MEXT, Japan.
Part of the calculations were performed on the JUGENE supercomputer at JSC.

\appendix

\section{}
\label{appendixB}
The use of random initial states has played a central role
in developing ``fast'' (i.e. ${\cal O}(D)$)
algorithms to compute the density of states and other similar quantities.
An early application of such an algorithm
to electron motion in disordered alloy models was given by Alben et al.~\cite{ALBE75}.
It was shown that the eigenvalue spectrum of a particle moving in continuum
space can be computed in the same manner~\cite{FEIT82}.
Fast algorithms of this kind proved useful to study various
aspects of localization of waves~\cite{RAED89,KAWA96,OHT97},
other one-particle problems~\cite{IITA97a,IITA97b,IITA99}
and many-body problems~\cite{VRIE93,HAMS00}.
The rigorous proof that this approach has remarkable
statistical properties, namely that
for large $D$ the statistical error vanishes as $1/\sqrt{D}$,
was given in Ref.~\cite{HAMS00}.

Following Ref.~\cite{HAMS00}, we consider real random variables $x_1$, $y_1$, $\ldots$, $x_D$, $y_D$, taking values in the interval $[-\infty, +\infty]$ and distributed according to the probability density
\begin{equation}
f(x_1,y_1,\ldots,x_D,y_D)=\frac{\Gamma(D)}{2\pi^D}\delta(x_1^2+\ldots+y_D^2-1),
\label{eqf}
\end{equation}
where $\Gamma(D)$ is the Gamma function.
Writing $c_n=x_n+iy_n$ for $n=1,\ldots,D$, we construct the random state
\begin{equation}
|\Phi\rangle=\sum_{n=1}^D c_n |\varphi_n\rangle,
\end{equation}
where $\{|\varphi_n\rangle\}$ is a complete set of orthonormal basis states of the $D$-dimensional Hilbert space,
which for the derivation that follows need not be specified further.

From Eq.~(\ref{eqf}), it directly follows that
\begin{equation}
\left<  c_k \right> =\int_{-\infty}^{+\infty} c_k f(x_1,\ldots,y_D)dx_1\dots dy_D=0,
\end{equation}
\begin{equation}
\left<  c_k^* c_{k'} \right> =\int_{-\infty}^{+\infty} c_k^* c_{k'} f(x_1,\ldots,y_D)dx_1\dots dy_D= \delta_{k,k'}D^{-1},
\end{equation}
and that $\left<  c_k c_{k'} \right>=0$.
It also follows from Eq.~(\ref{eqf}) that $|\Phi \rangle$
is a unit random vector with a uniform probability density on the hypersphere of dimension $D-1$.

Next we consider the projected state
\begin{equation}
|\Phi(\beta/2)\rangle\equiv e^{-\beta {\cal H}/2} |\Phi\rangle =\sum_{j=1}^D d_j e^{-\beta E_j /2} |E_j\rangle,
\label{eqProjectPhi}
\end{equation}
where $E_j$ ($|E_j\rangle$) denotes the $j$-th eigenvalue (eigenstate) of the Hamiltonian {\cal H} and
\begin{equation}
d_j =\sum_{j=1}^D R_{jn} c_n ,
\end{equation}
where $R_{jn}=\langle E_j|\varphi_n\rangle$ is the unitary transformation matrix.
The probability density of the random variables $d_j$ reads
\begin{equation}
f(d_1,\ldots,d_D)=\frac{\Gamma(D)}{2\pi^D}\delta(|d_1|^2+\dots+|d_D|^2 -1).
\label{eqfd}
\end{equation}
Hence the $d_j$ are distributed uniformly over the $D$-dimensional hypersphere.
Furthermore, we have $\left<d_j\right>=0$ and $\left<d_j^*d_{j'}\right>=D^{-1}\delta_{jj'}$.

Normalizing the state Eq.~(\ref{eqProjectPhi}) yields
\begin{equation}
|\Phi(\beta/2)\rangle =\sum_{j=1}^D \frac{d_j e^{-\beta E_j /2}}{\sqrt{\sum_{j=1}^D |d_j|^2 e^{-\beta E_j}}}
|E_j\rangle \equiv \sum_{j=1}^D a_j|E_j\rangle,
\label{eqNormProjectPhi}
\end{equation}
where
\begin{equation}
a_j=\frac{d_j p_j^{1/2}}{\sqrt{\sum_{j=1}^D |d_j|^2p_j}},
\end{equation}
and
\begin{equation}
p_j=\frac{e^{-\beta E_j}}{\sum_{j=1}^De^{-\beta E_j}},
\end{equation}
is the Boltzmann weight for the state $j$.
In general, the probability density of the coefficients $a_j$ is not uniform.

Next we want to show that for sufficiently large $D$,
we may replace $\sum_{j=1}^D |d_j|^2p_j$ by its average over the distribution Eq.~(\ref{eqfd}), that is by $D^{-1}$.
To prove this, we compute the average of
\begin{equation}
X^2 = \left( D^{-1}-\sum_{j=1}^D |d_j|^2p_j\right)^2,
\end{equation}
with respect to the distribution Eq.~(\ref{eqfd}).
We have
\begin{eqnarray}
\left< X^2\right> &=& D^{-2} - 2D^{-1}\sum_{j=1}^D p_j \left<|d_j|^2\right>
+\sum_{j=1}^D \sum_{j'=1}^Dp_j p_{j'}\left<|d_j|^2|d_{j'}|^2\right> \cr
&=&\sum_{j=1}^D \sum_{j'=1}^Dp_j p_{j'}\left<|d_j|^2|d_{j'}|^2\right> -D^{-2}.
\end{eqnarray}
Using Eq.~(2), (A12) and (A23) in Ref.~\cite{HAMS00}, we find
\begin{equation}
\left<|d_j|^2|d_{j'}|^2\right>=\frac{2\delta_{jj'}}{D(D+1)}+\frac{1-\delta_{jj'}}{D(D+1)}=\frac{1+\delta_{jj'}}{D(D+1)},
\end{equation}
yielding
\begin{eqnarray}
\left< X^2\right> &=& \frac{1}{D(D+1)}\sum_{j=1}^D p_j^2 - \frac{1}{D^2(D+1)} \cr
&\leq& \frac{D-1}{D^2 (D+1)}< \frac{1}{D^2}.
\end{eqnarray}
Invoking Markov's inequality~\cite{GRIM95}, it follows that
\begin{equation}
\mathbf{P}(X^2\geq D^{-1})<D^{-1}.
\end{equation}
In words, the probability that the error $X^2$ is larger than $D^{-1}$ is smaller than $D^{-1}$.
For the case at hand $D$ increases exponentially with the number of spins.
For instance for a system of $18$ spins and a random state with probability density Eq.~(\ref{eqf})
\begin{equation}
\mathbf{P}(X^2\geq 0.38\times 10^{-5})<0.38\times 10^{-5},
\end{equation}
suggesting that for all practical purposes, it is safe to assume that $X\approx 0$
and that with probability very close to one, the projected state Eq.~(\ref{eqNormProjectPhi})
can be written as
\begin{equation}
|\Phi(\beta/2)\rangle =D^{1/2} \sum_{j=1}^D d_j p_j^{1/2}|E_j\rangle.
\label{eqFinalProjectPhi}
\end{equation}
Putting $b_j=d_j p_j^{1/2}$, the probability density of random variables $b_j$ is
\begin{equation}
f(b_1,\ldots,b_D)=\delta(\frac{|b_1|^2}{p_1}+\dots+\frac{|b_D|^2}{p_D} -1)\frac{\Gamma(D)}{2\pi^D}\left( \prod_{j=1}^D p_j\right)^{-1/2}.
\label{eqfb}
\end{equation}

From these calculations, the following conclusions can be drawn about the projected state Eq.~(\ref{eqFinalProjectPhi}):
\begin{enumerate}
\item{
From Eq.~(\ref{eqFinalProjectPhi}) and the properties of random variables $\{d_j\}$, it follows directly that
on average
\begin{equation}
\langle\Phi(\beta/2)|Y|\Phi(\beta/2)\rangle=\frac{\mathbf{Tr}e^{-\beta {\cal H}}\; Y}{\mathbf{Tr}e^{-\beta {\cal H}}},
\end{equation}
assuming that $D$ is sufficiently large~\cite{HAMS00}.
Note that if $\langle\Phi|E_1\rangle\not=0$ ($|E_1\rangle$ denoting the non-degenerate eigenstate)
we have $\lim_{\beta\rightarrow\infty} \langle\Phi(\beta/2)|Y|\Phi(\beta/2)\rangle=\langle E_1|Y|E_1\rangle$,
independent of $D$.
}
\item{%
According to Eq.~(\ref{eqfb}), in general the projected state Eq.~(\ref{eqFinalProjectPhi}) is randomly distributed on
a hyper-ellipsoid, not on a hyper-sphere, in the Hilbert space.
This suggests that it may be of interest to extend the concept of canonical typicality
from states on an hyper-sphere to states on an hyper-ellipsoid
by introducing a non-zero inverse temperature beta.
}
\end{enumerate}

\begin{table}[t]
\caption{%
Same as Table~\ref{tab00} except that the composite system $S+E$ is prepared in a random state that
is typical for the composite system being in the canonical state at inverse temperature $\beta$.
}
\label{tab01}       
\begin{tabular}{lccccccccc}
\hline
\noalign{\smallskip}
$\beta$ & $b$ &$\sigma$ & $\delta$ & $S_b$ & $S_\beta$ & $S_{\tilde{\rho}}$ & $E_b$ & $E_\beta$ & $E_{\tilde{\rho}}$\\
\hline
&\multicolumn{8}{c}{$J=-0.5$, $\Delta=0.2$, $\Omega=0.5$}  &\\
\hline
1 & 0.976  & 0.004  & 0.002  & 2.671  &  2.666  &  2.671  & -0.202  &  -0.208  &  -0.202  \\
3	& 2.602	 & 0.025  &	0.016  & 2.042	&  1.838  &	 2.003  & -0.557  &  -0.630  &	-0.569  \\
6	& 3.744  & 0.045	&	0.077  & 1.462  &  0.633	&	 1.210  & -0.742  &  -0.920  &	 -0.799  \\
$\infty$ & 4.575 & 0.029	&	0.132	&	1.092  &  0.000  &	0.673  &	-0.832      & -1.000    &	-0.908\\
\noalign{\smallskip}
\hline
\end{tabular}
\end{table}

\begin{table}[t]
\caption{%
Same as Table~\ref{tab01} except that the coupling $\Delta=0.02$ instead of $\Delta=0.2$.
}
\label{tab03}       
\begin{tabular}{lccccccccc}
\hline
\noalign{\smallskip}
$\beta$ & $b$ &$\sigma$ & $\delta$ & $S_b$ & $S_\beta$ & $S_{\tilde{\rho}}$ & $E_b$ & $E_\beta$ & $E_{\tilde{\rho}}$\\
\hline
&\multicolumn{8}{c}{$J=-0.5$, $\Delta=0.02$, $\Omega=0.5$}  &\\
\hline
1	& 0.998 &	0.003 &	0.001	&	2.666	& 2.666 &	2.666	&	-0.207 & -0.208 &	-0.207 \\
6	& 5.981 &	0.032	& 0.020	&	0.637	& 0.633 &	0.573	&	-0.920 & -0.920 &	-0.928\\
\noalign{\smallskip}
\hline
\end{tabular}
\end{table}

\begin{figure}[t]
\begin{center}
\includegraphics[width=7cm]{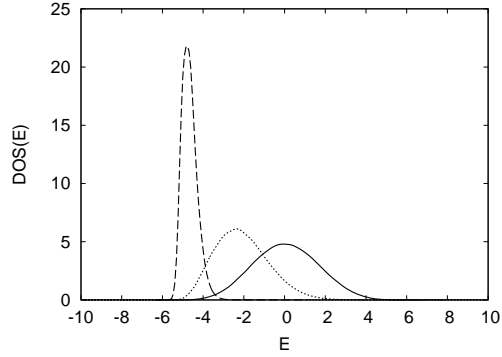}
\caption{%
Simulation results for the density of states (DOS) of the environment,
conditional on the initial state of the environment,
for the model parameters of Fig.~\ref{sfig4} corresponding to the weak-coupling case.
Solid line: $\beta=1$;
Dotted line: $\beta=3$;
Dashed line: $\beta=6$.
}
\label{sfig6}
\end{center}
\end{figure}

For completeness, we briefly discuss simulation results for the case
that we prepare the composite system $S+E$, not just the environment $E$, in a pure state that is typical for the canonical state
at a given $\beta$, namely
\begin{equation}
|\Psi(0)\rangle=\frac{ e^{-\beta H/2}|\Phi\rangle}{\langle\Phi|e^{-\beta H}|\Phi\rangle^{1/2}},
\label{inistate_case1}
\end{equation}
where $H$ is the Hamiltonian of the composite system, $|\Phi\rangle=\sum_j c_j |\phi_j\rangle$ is a pure state of the composite system,
$c_j$ is generated randomly according to the prescription given in Ref.~\cite{HAMS00}, and
$\{|\phi_j\rangle\}$ is an orthonormal set of basis states.
According to the theoretical analysis presented earlier in this Appendix,
averages taken with this pure state will yield the values that are equal to
those taken with respect to the canonical state with temperature $T=1/\beta$.

Solving the TDSE with the initial state Eq.~(\ref{inistate_case1})
yields results for $\sigma(t)$, $\delta(t)$, $\mathrm{var}(t)$ and $S(t)$ that, up to small fluctuations,
are constant in time, apparently consistent with the statement in Ref.~\cite{GOLD06},
that ``$\rho^{\Psi(t)}\approx \rho_\beta$ even at $t=0$ for typical wave functions''.
This may be taken as an indication that
the state Eq.~(\ref{inistate_case1}) is also ``typical'' in the sense
of ``canonical typicality''.
However, for the same reasons as those given earlier, this conclusion would be incorrect unless $\beta\rightarrow0$.


Obviously, the nanoscale models that we study are not ``thermodynamic'' in the usual sense,
but the number of states is quite large (Fig.~\ref{sfig6} shows the DOS of $\approx 4\times10^6$ states) and
it is possible, in principle, to find many (but not macroscopically many) states
with energies in a narrow interval such that the DOS in this interval is almost constant.
However, this is not sufficient to apply conventional statistical mechanics arguments:
What is required is that also the number of particles (22-35 in our work) is large.
Otherwise the fluctuations of the energy (which are proportional to the inverse square root
of the number of particles) are not small and the equivalence between microcanonical and canonical ensemble
is no longer guaranteed~\cite{KUBO85,GREI97}.

Our simulations show that the dependence of $\sigma(t)$, $\delta(t)$, $\mathrm{var}(t)$ and $S(t)$ on $\beta$
is qualitatively the same as in the case that we use the product state Eq.~(\ref{inistate}) as the initial state:
In all cases considered, the eigenvalues of the reduced density matrix converge to stationary values.
For comparison with Table~\ref{tab00}, in Tables~\ref{tab01} and \ref{tab03} we give the data extracted from the simulations
using Eq.~(\ref{inistate_case1}) as the initial state. It is clear that both Tables show the same qualitative features
as a function of $\beta$. For $\beta=1$ and $\Delta=0.02$ our results are in concert
with Tasaki's analytical results~\cite{TASA98} for weak coupling ($\beta\lambda\ll1$ in the notation of Ref.~\cite{TASA98}).

When we prepare the composite system in a pure state that is typical for the canonical state
at a given $\beta$, the simulation data of $\sigma(t)$, $\delta(t)$, $\mathrm{var}(t)$ and $S(t)$
show very little time-dependence (as discussed above).
However, as suggested by the data in Tables~\ref{tab01} and \ref{tab03},
the reduced density matrix is not equal to $e^{-\beta H_{S}}/\mathbf{Tr}_{S}e^{-\beta H_{S}}$
but is ``renormalized'' by the interaction $H_{SE}$ as can be seen by the perturbative
treatment that follows.

Up to the second order in the interaction Hamiltonian $H_{SE}$ we have~\cite{MORI08}
\begin{eqnarray}
e^{-\beta(H_S+H_E+H_{SE})}&=& e^{-\beta(H_S+H_E) }
-\int_0^\beta dx\;
e^{-(\beta-x)(H_S+H_E)} H_{SE}e^{-x ({ H}_{S}+{ H}_{E})}
\nonumber \\
&&+
\int_0^\beta dx \int_0^x dy\;
e^{-(\beta-x)({ H}_{S}+{ H}_{E})} { H}_{SE}e^{-(x-y)({ H}_{S}+{ H}_{E})} { H}_{SE}e^{-y ({ H}_{S}+{ H}_{E})}
\nonumber \\
&&+
{\cal O}(H_{SE}^3)
,
\end{eqnarray}
yielding $Z=Z_{S}Z_{E}(1-z_1+z_2)$
where $Z_{S}=\mathbf{Tr}_{S}e^{-\beta { H}_{S}}$, $Z_{E}=\mathbf{Tr}_{E}e^{-\beta { H}_{E}}$,
\begin{eqnarray}
z_1&=&\frac{1}{Z_{S}Z_{E}}\mathbf{Tr} \int_0^\beta dx e^{-(\beta-x)({ H}_{S}+{ H}_{E})} { H}_{SE}e^{-x ({ H}_{S}+{ H}_{E})}
\nonumber \\&=&
\frac{\beta}{Z_{S}Z_{E}}\mathbf{Tr} e^{-\beta({ H}_{S}+{ H}_{E})} { H}_{SE}
,
\end{eqnarray}
and
\begin{equation}
z_2=\frac{1}{Z_{S}Z_{E}}\mathbf{Tr} \int_0^\beta dx\int_0^x dy e^{-(\beta-x)({ H}_{S}+{ H}_{E})} { H}_{SE}e^{-(x-y)({ H}_{S}+{ H}_{E})}
{ H}_{SE}e^{-y ({ H}_{S}+{ H}_{E})}
,
\end{equation}
are the first and second-order correction, respectively.
Up to second order in the interaction Hamiltonian $H_{SE}$,
the reduced density matrix therefore reads
\begin{eqnarray}
\widehat\rho(\beta) &=&
\frac{\mathbf{Tr}_E e^{-\beta(H_S+H_E+H_{SE})}}{\mathbf{Tr}_S\mathbf{Tr}_E e^{-\beta(H_S+H_E+H_{SE})}}
= \frac{\mathbf{Tr}_E e^{-\beta(H_S+H_E+H_{SE})}}{Z_SZ_E(1-z_1+z_2)}
\nonumber \\
&=& \widehat \rho
\left(
1 + z_1 -z_2-\frac{1}{2} z_1^2
-\frac{1+z_1}{Z_E}\int_0^\beta dx\; \mathbf{Tr}_E e^{-\beta H_E} e^{x H_S}H_{SE} e^{-x H_S}
\right.
\nonumber \\
&&+
\left.
\int_0^\beta dx \int_0^x dy\;
\mathbf{Tr}_E e^{xH_S} e^{-(\beta-x+y)H_E} H_{SE}e^{-(x-y)(H_S+H_E)} H_{SE}e^{-y H_S}
\right)
\nonumber \\
&&+
{\cal O}(H_{SE}^3)
,
\end{eqnarray}
showing that in the thermal equilibrium state,
due to the interaction, we should expect a deviation from the canonical distribution of the system.

A direct numerical calculation of the various contributions is beyond our current capabilities and
we therefore leave this calculation for future research.
However, comparing the energy and entropy of the system in the canonical state
with the corresponding values obtained from the simulation of the composite system
provides some idea of how much the reduced density matrix
changes as a result of the interaction $H_{SE}$.
From Table~\ref{tab01}, we conclude that for $\Delta=0.2$
the differences $|S_\beta-S_{\tilde{\rho}}|$ and $|E_\beta-E_{\tilde{\rho}}|$
are of the order of 10\% or more, except for $\beta=1$.
Hence $\Delta=0.2$ definitely does not correspond to the case of weak interaction.
From Table~\ref{tab03}, it is clear that $\Delta=0.02$ and $\beta=1$ correspond to weak interaction
between environment and system because $S_\beta\approx S_{\tilde{\rho}}$ and $E_\beta\approx E_{\tilde{\rho}}$
but for $\beta=6$, $|S_\beta-S_{\tilde{\rho}}|$ is of the order of 10\%, hence not small.
Thus, we conclude that for both $\Delta=0.02$ and $\Delta=0.2$,
the effect of the interaction on the reduced density matrix is significant if $\beta=6$,
even if we prepare the composite system in a typical canonical state.

\bibliographystyle{jpsj}
\bibliography{/d/papers/epr}

\end{document}